\documentclass[twocolumn,showpacs,preprintnumbers,amsmath,amssymb]{revtex4}
\usepackage{graphicx}
\usepackage{dcolumn}
\usepackage{bm}

\begin{document}

\title{Vortex-antivortex wavefunction \\ of a degenerate quantum gas.}

\author{A.Yu.Okulov}
\email{okulov@kapella.gpi.ru}
\homepage{http://www.gpi.ru/~okulov}
\affiliation{
General Physics Institute of Russian Academy of Sciences 
Vavilova  str. 38, 119991, Moscow, Russia}

\date{\ March 1, 2009}

\begin{abstract}
A mechanism of a pinning of the quantized matter wave vortices 
by optical vortices in a specially arranged optical dipole 
traps is discussed.
The vortex-antivortex optical arrays of rectangular symmetry 
are shown to transfer
angular orbital momentum and form the "antiferromagnet"-like matter waves.
The separable Hamiltonian for matter waves in pancake trapping geometry
is proposed and 3D-wavefunction is factorized in a product of
wavefunctions of the 1D harmonic oscillator and 2D 
vortex-antivortex quantum state.
The 2D wavefunction's phase gradient
field associated via Madelung transform
with the field of classical velocities
forms labyrinth-like structure. The macroscopic quantum state composed of
periodically spaced counter-rotating
BEC superfluid vortices has zero
angular momentum and nonzero rotational energy.
\end{abstract}


\maketitle

\section{Introduction.}

Ultracold atomic gases \cite{Pitaevskii:2003,Pitaevskii:1999} have attracted
significant interest nowadays as a quantum
simulators of condensed matter systems
\cite{Lewenstein:2005} and as an
effective instrument for quantum information processing
\cite{Marangos:2005}.
The basic physical mechanisms for control of atomic
motion are magnetic trapping \cite{Phillips:1998}
and optical dipole trapping \cite{Ovchinnikov:2000}. Different
geometries of trapping fields were considered already, from the simplest one,
based upon 1D sinusoidal standing wave, formed by two counter-propagating
laser beams, to 3D artificial kagome potential landscape
built by specially arranged tilted laser beams configuration
\cite{Lewenstein:2005}. Such an artificial
potential lattices provide a rich opportunities for analog modeling of
many-body quantum systems e.g. Mott insulator transition \cite{Greiner:2002},
quantum Hall effect \cite{Cooper:1999,Ho:2001,Cornell:2004}, frustrated
quantum antiferromagnets
\cite{Lewenstein:2005}.
The antiferromanetic phase
is considered as an essential counterpart of
high temperature superconductivity
(HTSC) \cite{Maksimov:2007}. Recently the persistent currents in
toroidal "blue" detuned traps were reported \cite{Philips:2007}.
The pinning of the co-directed superfluid vortices in
different potential configurations
was analyzed with variational wavefunctions in order to
calculate vortex interaction energies
and it was shown that the 
most favorable allocations of vortices are at maxima of the lattice potential
\cite{Reijnders:2005}.
On the other hand the
elementary excitations of trapped ultracold gases,
namely abelian and nonabelian anyons were proposed as a promising
tool for error-tolerant quantum computing \cite{Kitaev:2002}. Noteworthy
the recent the mutual control of the
matter waves by light and vice versa demonstrated recently 
\cite{Hau:2002}.

	The interesting feature of an optical trapping by laser beams with wavefront
 dislocations is a possibility of guiding an atomic motion via non-potential
 optical fields i.e. by virtue of the optical vortices. Initially the
 elementary optical vortices like Laguerre-Gaussian beams (${\rm {LG}} $) were
 considered as toroidal traps for red detuned cold atoms \cite{Wright:2001}
 or hollow "tubes" under blue detuning from resonance \cite
{Letokhov:1990,Kuga:1997}. For a red detuned trap of toroidal
geometry \cite{Wright:2001} the optical torque had been predicted
 \cite{Allen:1994} which leads to angular acceleration of trapped atoms.
In the absence of optical torque
the macroscopic  quantum state of BEC in toroidal
trap had been studied by variational
approach and a wavefunction in the form of ${\rm {LG}} $
vortex spiral was obtained \cite{Abraham:2002}.
Classical dynamics of an atom trapped by helical
EM-fields guided by a nano-fiber
has been shown recently to exhibit spiral motion outwards the beam axis
as a result of optical torque \cite{Balykin:2006}.The formation and
acceleration of matter wave solitons
in toroidal quasi-1D ring trap
due to effect of an azimuthal oscillating electric field had been
studied \cite{Konotop:2007}.
The vortex-antivortex pairs in two-transverse dimensions
in non-rotating BEC traps of pancake geometry
were obtained numerically \cite{Perez:2003}.
The subject of the present article is an investigation of
the structure of a 
BEC wavefunction in $non-potential$ spatially periodic field
composed of overlapping optical vortices \cite{Okulov:2008}.
The paper is organized as follows. Section 2 describes the optical
dipole trap composed of overlapping $\rm LG$ vortices.
In section 3 the results
for conservative (potential or gradient) and dissipative (radiation pressure)
forces on two-level atom with electrical dipole transition
are summarized. Section 4 connects the classical tensors of electromagnetic
momentum and angular momentum with forces and torques 
on moving atom. In Section 5 the 
procedure of separation of variables for $\rm 3D$ Gross-Pitaevskii 
equation (GPE) is outlined along with numerical solution for $\rm 2D$ 
vortex-antivortex wavefunction. The procedure of separation of variables 
is closely connected to existence of different spatial scales
in GPE for pancake trap geometry,
namely longitudinal $\ell_{z}$, transversal $\ell_{\bot}$
and $healing$  length $\xi$.
The healing length $\xi$ appears as effective $nonlinear$ scale
related to cubic term in GPE \cite{Pitaevskii:1999} as a condition
of a balance between kinetic energy ("quantum pressure") and
interaction energy (two-body interaction):

\begin{figure}
\center{\includegraphics[width=0.95\linewidth] {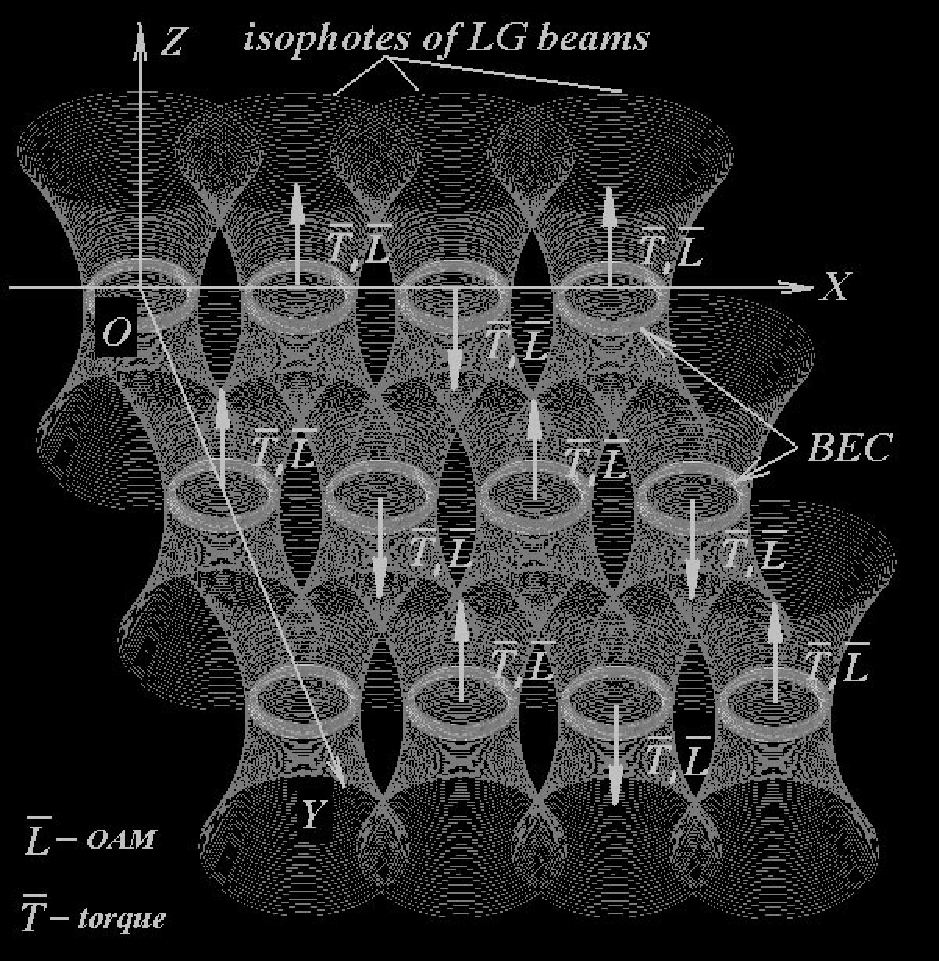}}
\caption{The isointensity lines for the superposition of
toroidal optical traps. 
Each elementary trap is $\bf LG $ beam propagating along  $Z$- axis.
The hyperboloidal surfaces are the loci of the maxima of intensity.
The rings at the $\bf LG $ beams bottlenecks are the isophotes which corresponds
to a maxima of a light intensity. Their diameter is chosen
close to experimentally observed in \cite{Chen:2001}, namely 
$d \approx 30 \mu m$,
wavelength $\lambda {\approx} 0.8 {\div} 1.06 \mu m$,
$\rm D \approx 180 \mu m$.}
\label{fig.1}
\end{figure}

\begin{equation}
\label{scales1}
 \xi = \sqrt{1 / 8 \pi  n a_s},
\end{equation}

where $n \approx |\Psi|^2$ is average density of a quantum
gas, $a_s$ is the $s$-wave
scattering length.
The $\ell_{z, \bot} = \sqrt{(\hbar / m \omega_{z, \bot})}$ are
often referred to as a 
characteristic widths (longitudinal and transversal respectively)
of the ground state of harmonic oscillator
for parabolic traps.
In pancake geometry considered below the following inequality is
valid \cite{Perez:1998,Kevrekidis:2008}:

\begin{equation}
\label{scales}
\ell_{z} < \ell_{\bot} < \xi.
\end{equation} 	

This allows to separate $z$ and $\vec r_{\bot}$
variables in $\rm GPE$ and factorize \cite{Konotop:2007,Manko:2004,
Okulov:2008} or in other terms
employ multiple scale expansion \cite{Perez:1998} to the
$\rm 3D$ wavefunction $\Psi$ to decouple it in
a product of longitudinal $\Psi_z$
and transversal $\Psi_{\bot}$ wavefunctions.
Section 6 devoted to estimation of
rotational energy and angular momentum of spatially periodic
macroscopic quantum state and section 7
summarizes the obtained results.

\section{Configuration of an optical labyrinth trap for neutral atoms.}

\begin{figure}
\center{\includegraphics[width=0.95\linewidth] {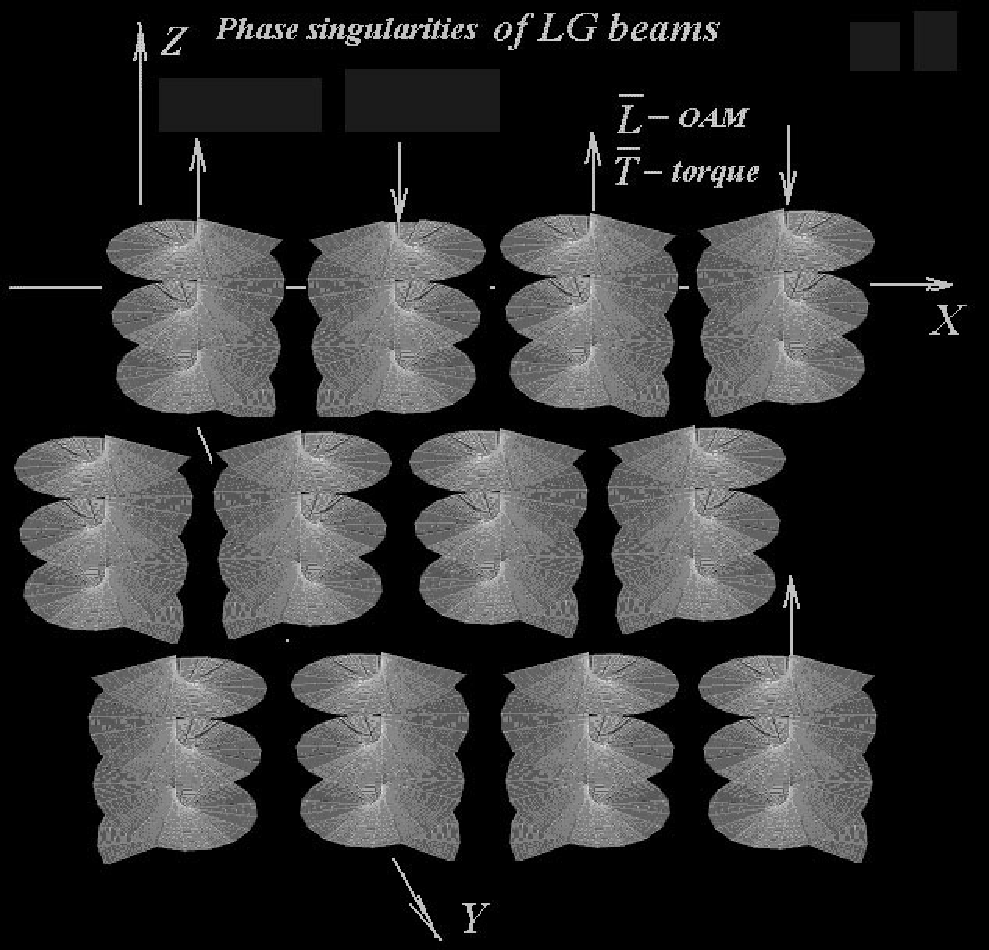}}
\caption{ The schematic representation of the periodically spaced optical
 vortices.
 Letters  $\vec L$ and $\vec T$  denote the angular momentum carried by
each phase singularity
and torque, respectively. The directions of $\vec T , \vec L$ are
shown to be parallel or anti-parallel
to $Z$ - axis.
The optical torques induce rotations of the cold atoms $clockwice$ or
$counter-clockwice$  respectively. The helical surfaces are
the snapshots of the phase of optical
field
at a given moment. The loci of helices are collocated with phase singularities.
The optical phase is undetermined along the axes of
helices denoted by arrows.
One round trip around the axis of a given helix in the $x,y$ - plane means
 the  $2 \pi$
change of the optical phase. The perfect match of helical wavefronts
between adjacent vortices is seen clearly. The trapped dipole moves
upstairs the helicoid, the radius of rotation is gradually increased
due to torque. The passage from one helix to another is possible due
to perfect match of adjacent helical surfaces.}
\label{fig.2}
\end{figure}

	The rectangular optical vortex lattices are spontaneously formed in diode-pumped
 microchip lasers with slightly focusing output coupler in a wide range of
 experimental parameters \cite{Chen:2001}. The optical patterns observed in this
 experiment are nonlinear eigenmodes of Fabry-Perot resonator with sufficiently
large Fresnel number$N_{fr}={\frac {k D^2}{L_r}}$, ranging
from $\it 100$ to $\it 1000$: ,  where $k=2 {\pi}/{\lambda}$ , $L_r$  is optical  length of cavity, $D$ is
diameter of optically pumped area inside host
crystal which is approximately equal to
diameter of generated optical array. It is convenient
to approximate the laser eigenmode obtained numerically \cite{Staliunas:1995}
and experimentally \cite{Chen:2001} as a superposition
of co-propagating and overlapping $\rm LG$'s with
the unit topological charges \cite{Okulov:2008}:

\begin{eqnarray}
\label{hold_vort}
\ E(\vec{r},z=0 ) \approx  E_0 {\:}
{  \exp{[- {\:} 
\frac {{| \vec{r} |}^2}{D^2}}]}{\:}
\sum_{jx,jy}(-1)^{\it jx+jy}
{| \vec{r}-\vec{r}_{jx,jy} |}
& &  \nonumber \\
{\times {\:}\exp{{\:}[- {\:}\frac 
{{| \vec{r}-{\vec{r}_{\it jx,jy}} |}
^{{\:}2}}{d^2}}}+
{\:}{i}{\:}
\ell_{EM}{\:} {\it Arg}(\vec{r}-\vec{r}_{jx,jy} ){\:}],{\:}
{\:}{\:}
\end{eqnarray}

where  $jx,jy$  are integer indices for positions of elementary $ \rm {LG}$
 vortices spanned with period $p$ in $(z=0,x,y)$ - plane, $d$ is
 diameter of the bottleneck of $\rm LG$ and $\ell_{EM}$ is topological charge
of elementary $ \rm {LG}$ optical vortex,
$\vec r=(x,y)$  is a vector in transverse plane,
$\vec r_{jx,jy}$  is a vector indicating positions of elementary vortices.
The resulting interference pattern obtained via superposition
(\ref{hold_vort})  is ordered in
"antiferromagnet"-like
lattice with angular momenta alternating from one site to another(fig.1)
 \cite{Okulov:2004}.
Such 2D periodic optical vortex array forms a superposition of multiply
connected
toroidal optical traps (fig.  \ref{fig.2}). The motion
of cold atoms is controlled by
 combined
action of optical dipole force \cite{Ovchinnikov:2000}
and radiation pressure force \cite{Allen:1994,Balykin:2006}.
Alternative useful approximation for optical field is
is a superposition of the several major 
 Fourier components \cite{Cornell:2006}
using e.g. exact formula for free space propagation 
of periodical optical field
 from \cite{Okulov:1990}:

\begin{eqnarray}
\label{tetraedral}
\ E(\vec{r},z ) \approx  E_0 {\:}
{  \exp{[ikz - {\:} 
\frac {{| \vec{r} |}^2}{D^2}}]}{\:}
{\:} \sum_{jx ,jy} A_{jx ,jy}{\:} 
& &  \nonumber \\
\exp{\:}
[ {\:}i2 {\pi} \lbrace  \frac {x {\cdot} jx} {p } +\frac {y \cdot  jy} {p }+\frac { z} {2k} (\frac {jx ^2} {p^2 }+ \frac {jy^2} {p^2 }) \rbrace ].
{\:}{\:}{\:}
\end{eqnarray}

It is worth to mention here that a similar geometry of the optical array
produced with the programmable spatial light
modulators \cite{Grangier:2004} and microlens arrays \cite{Ertmer:2002}
giving the possibility of fine tuning of phases $\phi_{jx,jy}$ and amplitudes
 $Amp({\vec{r}-\vec{r}_{jx,jy}})$ of a given beam in array:

\begin{eqnarray}
\label{adjust_vort_array}
\ E(\vec{r},z=0 ) \approx  E_0 \sum_{jx,jy}{Amp{\:}[{\:} {(\vec{r}-\vec{r}_{jx,jy}) }{\:}}/d^2]
& &  \nonumber \\ {\:} {\exp{\:}[ {\:}{{i}}{\:} \ell_{EM} Arg( \vec{r}-\vec{r}_{jx,jy} ){\:}+ {\:}{{i}}{\:} \phi_{jx,jy}}],
\end{eqnarray}

\section{Conservative and dissipative forces on moving neutral atoms.}
	
The knowledge of a particular spatial distribution of $\vec E({z,\vec r})$
permits the explicit calculation of the expectation
values of $\hat {\vec {F}}$ i.e. classical force $\vec F$
on resonant atom
with an electrical dipole transition.
For  conservative (or reactive)
part $\vec F_{R}$ we have expression for gradient force \cite{Ovchinnikov:2000}:

\begin{equation}
\label{force_reactive}
 {\langle \vec {F} \rangle}_{R} = - \nabla \lbrace
{{\langle {\hat {\vec d}}\cdot {\vec E({z,\vec r})} \rangle}}\rbrace.
\end{equation} 	

For dissipative component \cite{Allen:1994} the formula
for radiation pressure follows:

\begin{eqnarray}
\label{force_whole}
{\langle \vec {F} \rangle} ={\langle \vec {F} \rangle}_{R} +
{\langle \vec {F} \rangle}_{D}={\frac {d \langle{\vec P_{at}}\rangle}{dt}}={\:}{\:}{\:}{\:}{\:}{\:}
& & \nonumber \\
 {\frac {d}{dt}}{\lbrace \langle {\vec P_{at}(t=0)}+
{\frac {i}{\hbar}}
{\int\limits_{0}^{t}}[\hat H(t^{'}),{\vec P_{at}(t^{'})}] d{t^{'}}
\rangle \rbrace}=
& &\nonumber \\  i{\hbar}{\lbrace n(2 n_{e}-1)+n_e\rbrace}
[f^{*}{\nabla f}I_{S}(t)-f{\nabla f^{*}}{I_{S}^{*}}(t)],{\:}{\:}{\:}{\:}
\end{eqnarray}
where $\hat {\vec {P_{at}}}$ is an atomic momentum,
$\hat H(t^{'})$ is a Hamiltonian of the atom in rotating wave approximation,
$n_e=\langle \Psi |{\hat {n_e}}| \Psi \rangle $ is an average number of
atoms on upper level,
$n=\langle \Psi |{ {{\hat a}^{+}_0} \hat {a_0}}| \Psi \rangle $ is the average 
 number of photons in a given electromagnetic mode (${{\rm {LG}}_{01}} $
 mode in our case),

 \begin{eqnarray}
\label{defin}
f  \cong {\vec D_{12}}\cdot {\vec E_(z, {\vec r})}
= G(z, \vec r) exp(i \Theta(z, \vec r)) ;
& & \nonumber \\
{\:}{\:}{\:}{\:}{\:}{\:}I_{S}(t)={\int\limits_{0}^{t}}exp{\:}[i \Delta t^{'}/2]
{\frac {sin(\Delta t^{'} /2)}{\Delta /2}},
\end{eqnarray} 	

where $\vec D_{12}$ is electric dipole matrix element of the two-level
 transition,
${ \Delta}= \omega_0 - \omega + \delta $ is a detuning and $\delta$ is
given by:
 \begin{equation}
\label{recoil}
\delta = {\frac {1}{2M}} {\Bigl[} \frac {{\langle \bf P \rangle}\cdot
\nabla f
+ {\nabla f\cdot \langle \bf P \rangle} }{f} \Bigr]_0.
\end{equation}

Fortunately the back action of the radiation scattered by
freely moving atom upon incident field is negligibly small for currently
achieved atom densities in optical dipole traps and a field
${\vec E_({\vec r},z) }$ could be substituted from
classical solution of Maxwell equations. As a result
an $azimuthal$ component
of semiclassical dissipative 
force ${\langle F_{\phi} \rangle}_D$
on atom in $\rm LG$ helical optical beam
is as follows \cite{Allen:1994}:

\begin{equation}
\label{force_diss_azim}
 {\langle \ F_{\phi} \rangle}_D =
{\frac {2 \hbar  {\Gamma}{{\Omega}^2}_{kpl}
({\vec r},z) }{{\Delta}^2+2{{\Omega}^2}_{kpl}({\vec r},z)  +\Gamma^2}}
{\frac {\ell_{em}}{r}}{\vec \phi},
\end{equation}
where
${{\Omega}^2}_{kpl}({\vec r},z) = {\vec D}_{12} \cdot {\vec E}(z,
 {\vec r})/{\hbar}$, $\Gamma$ is atomic linewidth.
This component is responsible for the angular acceleration of the 
atomic dipole around
$Z$ - axis. Two other components of dissipative force were also
obtained in explicit form.
The longitudinal component of the radiation pressure force is:

\begin{equation}
\label{z_force_diss}
 {\langle \ F_{z} \rangle}_D =
{\frac { \hbar k {\Gamma} I({\vec r},z)
}{1+I({\vec r},z) +\Delta^2 / \Gamma^2}},
\end{equation}

and radial force ${{\langle {{F_{r}}} \rangle}_D}$ is :

\begin{equation}
\label{radial_force_diss}
 {\langle  F_{r} \rangle}_D =
{\frac { \hbar k {\Gamma} \nabla I({\vec r},z) {\vec r}
}{2 \epsilon_0 c}}
{\Bigl[}
{\frac { 1}{1+I({\vec r},z) +\Delta^2 / \Gamma^2}}
{\Bigr]},
\end{equation}
\newpage

As a result complete $3D$ classical motion
of the atom with resonant dipole transition having mass $m$
in isolated optical vortex 
is governed by following equations \cite{Balykin:2006}:

\begin{equation}
\label{two_newton}
 m \ddot{z} =F_{z} {\:}{\:}{\:}; {\:}{\:}{\:}{\:}{\:}{\:}
m \ddot{r} =F_{r}+m r {\dot{\phi}}^2 {\:}{\:}{\:} ;{\:}{\:}{\:}{\:}{\:}{\:}
m r \ddot{\phi} =-2 m \dot{r} \dot{\phi}+F_{z}.
{\:}{\:}{\:}{\:}{\:}{\:}
\end{equation}
On the other hand in vector notations we have the following 
equation for classical motion
of a particle with a mass $m$:

\begin{equation}
\label{vector_newton}
 \dot{\vec L} ={\frac {d}{dt}} {\lbrace m {\vec r}^{{\:} 2} {\vec \omega} \rbrace}= {\vec T(\vec r)}=[{\vec r} \times
{\vec F}]
{\:}{\:}{\:}{\:}{\:}{\:}
\end{equation}

The straitforward generalization of this equation
for rectangular array of equidistantly spanned superimposed
vortices reads:

\begin{eqnarray}
\label{newton_array}
 m \ddot{\vec r} =\sum\limits_{jx ,jy} \vec {F}
({\vec r}-{\vec r_{jx ,jy}},z)
{\:}; {\:}
& &\nonumber \\
 m \ddot{\vec r} =\sum\limits_{jx \pm 1 ,jy \pm 1} 
\vec {F}({\vec r}-
{\vec r_{jx ,jy}},z),
\end{eqnarray}
where second equation in  (\ref{newton_array}) takes into account
only nearest neighboring overlapping optical vortices.		 		 	
The azymuthal components of
 the Pointing
vector accelerates the condensate around vortex axis  (fig.  \ref{fig.2}
 Z-axis) in the following way.
The classical trajectories initially located near vortex core with velocities
 close to zero are
almost circular. The radius of rotation is gradually increased until atom
 would reach the separatrix,
analogously to \cite{Balykin:2006}.
Then atom passes to another basin of atraction located around ajaicent vortex
 core\cite{Okulov:2008}.The numerical solution of  (\ref{newton_array}) shows how classical dipole
 trapped in optical
labyrinth field moves along Mobius-like trajectories around zeros of intensity,
roaming from one phase singularity to another.
 
\section{Densities of the linear and angular electromagnetic momenta}

\begin{figure}
\center{\includegraphics[width=0.8\linewidth] {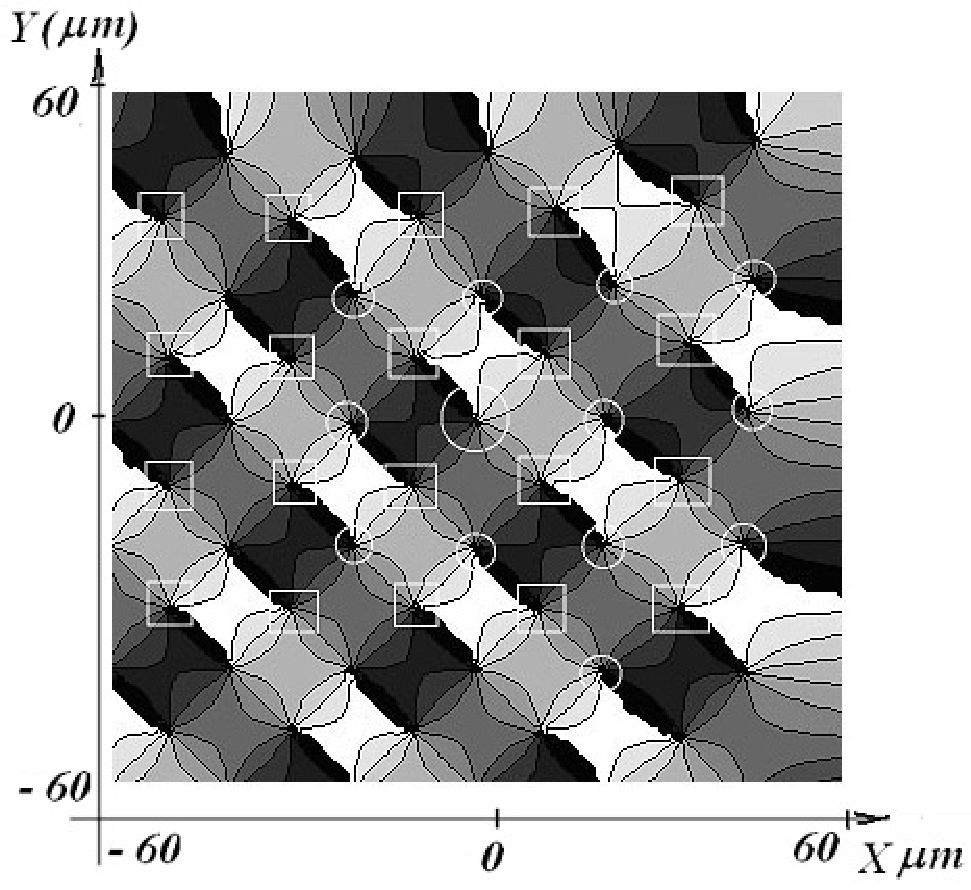}}
\caption{ The $\bf 2D$ plot of the phase of
electric field $Arg[E({\vec r},z=0)]$,
eq. (\ref{hold_vort}) in the  $x,y$- plane.
The locations of the elementary optical vortices with positive topological
charges $\ell = +1$  are labeled by white circles. The surrounding vortices
 having
the opposite charges  $\ell = -1$  are labeled by white squares. Such
flip-flop distribution of local angular momentum arises
due to interference of the overlapping LG beams in eq. (1). }
\label{fig.3}
\end{figure}

Using effective cross-section $\sigma_{opt}$ of an atom which 
scatters the optical 
field $\vec E({\vec r},t)$ ,
the equations  (\ref{force_diss_azim}-\ref{radial_force_diss}) could be
reformulated in a terms of electromagnetic
energy flux (Pointing vector) $\vec S=\epsilon_0 c^2 [\vec E \vec B]$, momentum density $\vec P({\vec r},t)$,
momentum flux density  $T^{jl}$,  angular momentum
density $\vec M({\vec r},t)$, angular momentum flux density  $M^{jl}$
\cite{Barnett:2002}.
Classically the force $d \vec F$ experienced by infinitesimal element of
 surface of a physical body
in electromagnetic field could be evaluated by multiplying of the
 electromagnetic pressure $p$
by an infinitesimally small surface element $ds$  having local
normal $\vec n (\vec R)$  :
\begin{eqnarray}
\label{em_pressure}
d \vec F \approx p {\:} \vec n {\:}(\vec R) {\:}ds ,
\end{eqnarray}
The physical meaning of expression (\ref{em_pressure}) is momentum flux through
infinitesimal surface element  $ds$ of  area per a unit time. Taking into
account only normal component of optical flux i.e.
component parallel to normal $\vec n$ the  force  $d \vec F$ on this particular
surface element in tensor notations is as follows:
  $ d F_j = T^{jl}{\:} ds_{l}$ ,

where $T^{jl}=\frac{\delta_{jl}}{2} [{\epsilon_0}
{|{\vec E}|}^2+{\mu_0}^{-1} {|{\vec B}|}^2] -{\epsilon_0} E_j E_l -
 {\mu_0}^{-1} B_j B_l $ is momentum flux density \cite{Barnett:2002},
 ${\delta_{jl}}$  is Kronecker's delta.
The magnitudes of $T^{jl}$ components define the magnitudes
 and directions of optical forces on atoms in
the vicinity of phase singularity  (fig.  \ref{fig.2}).
The components of the energy flux density (Pointing vector)
$\vec S=\epsilon_0 c^2 [\vec E \times \vec B]$
are proportional to linear momentum density
$\vec P= {\vec S}/{c^2} =\epsilon_0 [\vec E \times \vec B]$
components:

\begin{eqnarray}
\label{momentum_flux}
P_z =2 {\:}\epsilon_0{\:} c  {|\vec E({\vec r,z})|}^2 {\:}
{\:}{\:}{\:}{\:}{\:} ;{\:}{\:}{\:}{\:}{\:}{\:}
P_r =\epsilon_0{\:} {\frac {\omega k r z}{z^2+{z_R}^2}} 
{|\vec E({\vec r,z})|}^2
& &\nonumber \\ P_{\phi} =\epsilon_0{\:} [ {\frac {\omega \ell_{em}}{r}} {|\vec E({\vec r,z})|}^2 -
 {\frac {\omega \sigma } {2}}
{\frac {\partial {|\vec E({\vec r,z})|}^2 } {\partial r }}],
{\:}{\:}{\:}{\:}{\:}{\:}
\end{eqnarray}

where $ \sigma$ is a light polarization equal to $\it {0, \pm 1}$ for plane and circular
polarizations respectively \cite{Allen:1992}. 
The major, i.e. longitudinal component  $P_z$ (\ref{momentum_flux}) is 
responsible for optical pressure force. 
The radial component  $P_r$  (\ref{momentum_flux}) pushes atomic dipole 
outwards the beam axis ($Z$) and the last, 
azimuthal component   $P_{\phi}$ of  (\ref{momentum_flux}) accelerates atom 
around $Z$  - axis.  Angular momentum 
density $\vec M$  is defined as a vector product, analogously to definition of 
mechanical torque $\vec T=[\vec r \times \vec F]$:

\begin{eqnarray}
\label{angular_momentum_density}
\vec M=[\vec r {\times} \vec P({\vec r,z})]={\frac {[\vec r {\times}
 \vec S({\vec r,z})]}{c^2}} =
{\:}\epsilon_0{\:} {\vec r} {\times}{\:}{\:}{\:}{\:}{\:}{\:}
& &\nonumber \\
{[\vec E({\vec r,z}) \times \vec B({\vec r,z})]}=
-{\frac {\ell_{em} z }
{\omega r}} {|\vec E({\vec r,z})|}^2 {\vec r}-{\:}{\:}{\:}{\:}
& &\nonumber \\
{\frac {r}{c}}[{\frac {z ^2 }
{ (z^2 - {{z_R}^2 })}}-1]
{|\vec E({\vec r,z})|}^2 {\vec \phi}
+{\frac {\ell_{em} {|\vec E({\vec r,z})|}^2 }{\omega z}}
{\vec z},{\:}{\:}{\:}{\:}{\:}{\:}{\:}
\end{eqnarray}

where $z_R=k D^2$ is Rayleigh range. The field of transversal 
(in the plane $x,y$) momentum 
density $\vec P(x,y)$  is proportional 
to the phase gradient (fig.\ref{fig.3}):

\begin{equation}
\label{em_phase_grad}
\vec P(x,y) \approx \nabla \lbrace Arg [{\vec E(x,y,z=0)}] \rbrace
{\:}{\:}{\:}{\:}{\:}{\:}
\end{equation}
The interference between adjacent LG - beams generates additional
optical vortex lattice with opposite angular
momenta \cite{Okulov:2008,Okulov:2004}. Under conditions of the far
detuning from resonance it is possible to construct the optical
dipole potential. This potential is separable
in the geometry considered in
\cite{Okulov:2008} (Fig. \ref{fig.1}):

\begin{eqnarray}
\label{TRAPOP1}
\ { V_{ext}(\vec{r_{\bot}},z)}=V_z +V_{_{\bot}} = 
{\frac { m{\:} {\omega{_z}}^2 {z^2}}{2 }}-
& &\nonumber \\
Re[{\alpha(\omega)] {\:}
} |{\vec E}(\vec{r}_{\bot} ) |{\:}^2
+ {\frac { m{\:} {\omega{_{\bot}}}^2 { | (\vec{r}_{\bot} ) |
{\:}^2  } }{2} }, {\:}{\:}{\:}{\:}{\:}{\:}
\end{eqnarray}

where $\omega_{z}, \omega_{\bot}$ are the transversal and longitudinal
frequencies of optical trap respectively\cite{Pitaevskii:2003}. 
$\alpha(\omega)$ is the polarizability of atom\cite{Ovchinnikov:2000}:

\begin{equation}
\label{polar}
\alpha(\omega) = 6 \pi \epsilon_0 {\:} c^3  {\frac {\Gamma / {\:}{\omega_0}^2
 {\:} } {(\omega_0^2-\omega^2-{{i}}(\omega^3 / {\omega_0}^2)\Gamma)}},
\end{equation}
which is real for large detuning from resonance $\omega - \omega_0$.

\section{Gross-Pitaevskii equation with separable Hamiltonian.}

We solve the Gross-Pitaevskii equation ($\rm GPE$) for macroscopic BEC
 wavefunction\cite{Pitaevskii:2003,Pitaevskii:1999}:

\begin{equation}
\label{GPE5}
 {\:}{\:}{\:}{\:}{\:}{\:}
\ {{{i}} \hbar}{\:}{\frac {\partial {\Psi}(\vec{r},t )}
{\partial t}} = \hat H  {\Psi}(\vec{r},t ), {\:}{\:}{\:}{\:}{\:}{\:}
\end{equation}
with following separable Hamiltonian  \cite{Perez:1998,Manko:2004,Kevrekidis:2008}:

\begin{eqnarray}
\label{hamilt_separ}
 \hat H =  \hat H_{\bot} +  \hat H_{||}= 
-{\frac {\hbar^2}{2 m}} \Delta_{\bot}
-{\frac {\hbar^2}{2 m}} {\frac {\partial^2} 
{\partial z^2}} + {\frac { m{\:} {\omega{_z}}^2 
{z^2}}{2 }}{\:}- 
& &  \nonumber \\
Re[{\alpha(\omega)] {\:}
} |{E}(\vec{r}_{\bot} ) |{\:}^2+
{\frac { m{\:} {\omega{_{\bot}}}^2 { | (\vec{r}_{\bot} ) |
{\:}^2  } }{2} } + 
& &  \nonumber \\
{\frac{4{\pi}
{\hbar}{\:}^2{\:}{a_s({\vec{B}})}}{m}}
|{\Psi}(\vec{r},t ) |{\:}^2, 
\end{eqnarray}

where $a_s$ is $s-wave$ scattering length.
For the asymmetrical optical trap when $\omega_{\bot} < \omega_z$
 and when "healing length" $\xi = (8{\:}\pi{\:} n
{\:}a_s )^{-1/2}$ \cite{Pitaevskii:1999}
 is larger than longitudinal
harmonic oscillator length  $ \sqrt{\hbar / m \omega_{z}}$
\cite{Perez:1998,Kevrekidis:2008}. Hence 
it is reasonable to seek the solution for the eq.(\ref{GPE5})
with Hamiltonian (\ref{hamilt_separ}) by method of separation of variables \cite{Manko:2004}.
The substitution of the factorized wavefunction 
$\Psi(z,r_{\bot},t) = \Psi_{\bot}(r_{\bot},t) \Psi_{||}(z,t)$ 
in (\ref{GPE5}) gives:

\begin{eqnarray}
\label{GPE_subst}
\  {{ {i}} \hbar}{\:} {\Bigr [} {\Psi_{||}} {\frac {\partial {\Psi_{\bot}}}
{\partial t}} +{\Psi_{\bot}}{\frac {\partial {\Psi_{||}}}{\partial t}}
{\Bigl ]} =
 -{\Psi_{||}}{\frac {\hbar^2}{2 m}} \Delta_{\bot} 
{\Psi_{\bot}}-
& &\nonumber \\
Re[{\alpha(\omega)] {\:}} |{E}(\vec{r}_{\bot} ) |{\:}^2 {\Psi_{||}}
{\Psi_{\bot}}
-{\frac {\hbar^2}{2 m}} {\frac {\partial^2 {\Psi_{||}}}{\partial z^2}}
{\Psi_{\bot}}+ {\frac { m{\:} {\omega{_z}}^2 {z^2}}{2 }}{\Psi_{||}}{\Psi_{\bot}}
&&\nonumber \\
+ {\frac { m{\:} {\omega{_{\bot}}}^2 { | (\vec{r}_{\bot} ) |
{\:}^2  } }{2} }{\Psi_{||}}
{\Psi_{\bot}}+
{\frac{4{\pi}{\hbar}{\:}^2{\:} {a_s( {\vec{B}})}}{m}} {\Psi_{\bot}}
|{\Psi_{\bot}} |{\:}^2 {\Psi_{||}} |{\Psi_{||}} |^2.
{\:}{\:}
\end{eqnarray}

As a consequence of the different spatial scales $\ell_z < \ell_{\bot}< \xi$
the starting GPE (\ref{GPE5}) is exactly decoupled in a
way analogous to \cite{Perez:1998, Manko:2004}:

\begin{equation}
\label{osc1d}
\ {{ {i}} \hbar}{\:}{\Psi_{\bot}}{\frac {\partial {\Psi_{||}}}{\partial t}} =
 -{\Psi_{\bot}} {\frac {\hbar^2}{2 m}} {\frac {\partial^2 {\Psi_{||}}}
{\partial z^2}}+ {\Psi_{\bot}} {\frac { m{\:} {\omega{_z}}^2 {z^2}}{2 }}
{\Psi_{||}}
\end{equation}
and
\begin{eqnarray}
\label{GPE_subst1}
\  {i} \hbar{\:}{\Psi_{||}} {\frac {\partial {\Psi_{\bot}}}{\partial t}}  =
 -{\Psi_{||}}{\frac {\hbar^2}{2 m}} \Delta_{\bot} {\Psi_{\bot}}-
Re[{\alpha(\omega)] {\:}} |{E}(\vec{r}_{\bot} ) |{\:}^2 {\Psi_{||}}
{\Psi_{\bot}}
&+&\nonumber \\
{\frac { m{\:} {\omega{_{\bot}}}^2 { | (\vec{r}_{\bot} ) |{\:}^2  } }{2} }{\Psi_{||}}
{\Psi_{\bot}}
+{\frac{4{\pi}{\hbar}{\:}^2{\:} {a_s( {\vec{B}})}}{m}} {\Psi_{\bot}}
|{\Psi_{\bot}} |{\:}^2 {\Psi_{||}} |{\Psi_{||}} |{\:}^2.
\end{eqnarray}

The solution of (\ref{osc1d}) for ground state 
inside longitudinal
parabolic trap (harmonic oscillator)
becomes evident:

\begin{equation}
\label{ground1D}
\ {\Psi_{||}}={{ ( \frac {m \omega_z }{ \pi \hbar})^{1/4} }}{\:}exp{\:}
[-m \omega_z z^2/ (2 \hbar)- {{ {i}}{\:}\omega_{z}  {\:}t}]
\end{equation}

Next, by virtue of multiplying (\ref{GPE_subst1})  by complex
conjugate $\Psi_{||}$ , integrating it by $z$
from $ - \infty$ to  $ \infty$, i.e. 
using normalization conditions:

\begin{equation}
\label{normapsi}
\int^\infty_{-\infty}|{\Psi_{||}}(z,t ) |{\:}^4 d z= 1/2 {\:}{\:}{\:}{\:}
and {\:}{\:}\int^\infty_{-\infty}|{\Psi_{||}}(z,t ) |{\:}^2 d z= 1
\end{equation}

one obtains:

\begin{eqnarray}
\label{GP2D}
\  {{ {i}} \hbar}{\:}{\frac {\partial {\Psi_{\bot}}}
{\partial t}} =
 -{\frac {\hbar^2}{2 m}} \Delta_{\bot} {\Psi_{\bot}}+
Re[{\alpha(\omega)]
{\:}} |{E}(\vec{r}_{\bot} ) |{\:}^2 {\Psi_{\bot}}
&+&\nonumber \\
{\frac{4{\pi}{\hbar}{\:}^2{\:}{a_s({\vec{B}})}}{m}}
{\Psi_{\bot}}
|{\Psi_{\bot}} |{\:}^2
\end{eqnarray}

Next the transversal component of wavefunction is obtained numerically 
via split-step FFT algorithm on $512 \times 512$  points square computational 
mesh \cite{Okulov:2008}. 
In order to emulate the optical torque not included yet in our
computational model
we prepared a special initial conditions for the transversal wavefunction
$\Psi_{\bot}(t=0,\vec r_{\bot})$ in the form of the rectangular 
array in the form of (\ref{tetraedral}) and found the convergence
with reasonable
accuracy ($\approx 10^{-3}$) after $200  {\div} 500 $ iterates. It is worth to mention specially, that
the rectangular symmetry of numerical solution
of 2D GPE (\ref{GP2D}) is imposed by combined action
of initial trial wavefunction  $\Psi^{0}(\vec r, t)$ having
rectangular symmetry similar to  trapping potential
 and weakness of cubic term in (\ref{GP2D}) compared to trapping term.
The initial guess for iterative explicit split-step FFT method
was choosen in the form 2D vortex
lattices  (\ref{tetraedral}). After $n_i=20 \div 150$ iterates
the  $\Psi^{n_i}(\vec r, t)$  remained well correlated with
trapping field  ${E}(\vec{r} ) $. Next within
following $n_i=200 \div 500$   iterates the amplitude of wavefunction 
decreased down to $10 \div 100$ times smaller than initial
guess amplitude, due to the intrinsic dissipation 
of the numerical method, which uses the spatial filtering of the 
Fourrier components of 
high spatial frequencies \cite{Okulov:1993,Okulov:1994}.

It is well known that spatially periodic optical trapping leads to Bloch
waves and gaps in BEC energy spectrum. The gaps and cubic nonlinearity
affect each other. In order to simplify numerical solution of (\ref{GP2D}) the
parameters of equation were adjusted in such a way, that the last two terms in (\ref{GP2D}) 
would almost exactly cancel each other:

\begin{equation}
\label{GP_comp}
Re[{\alpha(\omega)] {\:}} |{E}(\vec{r}_{\bot} ) |{\:}^2
 {\Psi_{\bot}}+
{\frac{2{\pi}{\hbar}{\:}^2{\:}{a_s({|\vec{B}|})}}{m}}{\Psi_{\bot}}
|{\Psi_{\bot}} |{\:}^2 \approx 0
\end{equation}

\begin{figure}
\center{\includegraphics[width=0.8\linewidth] {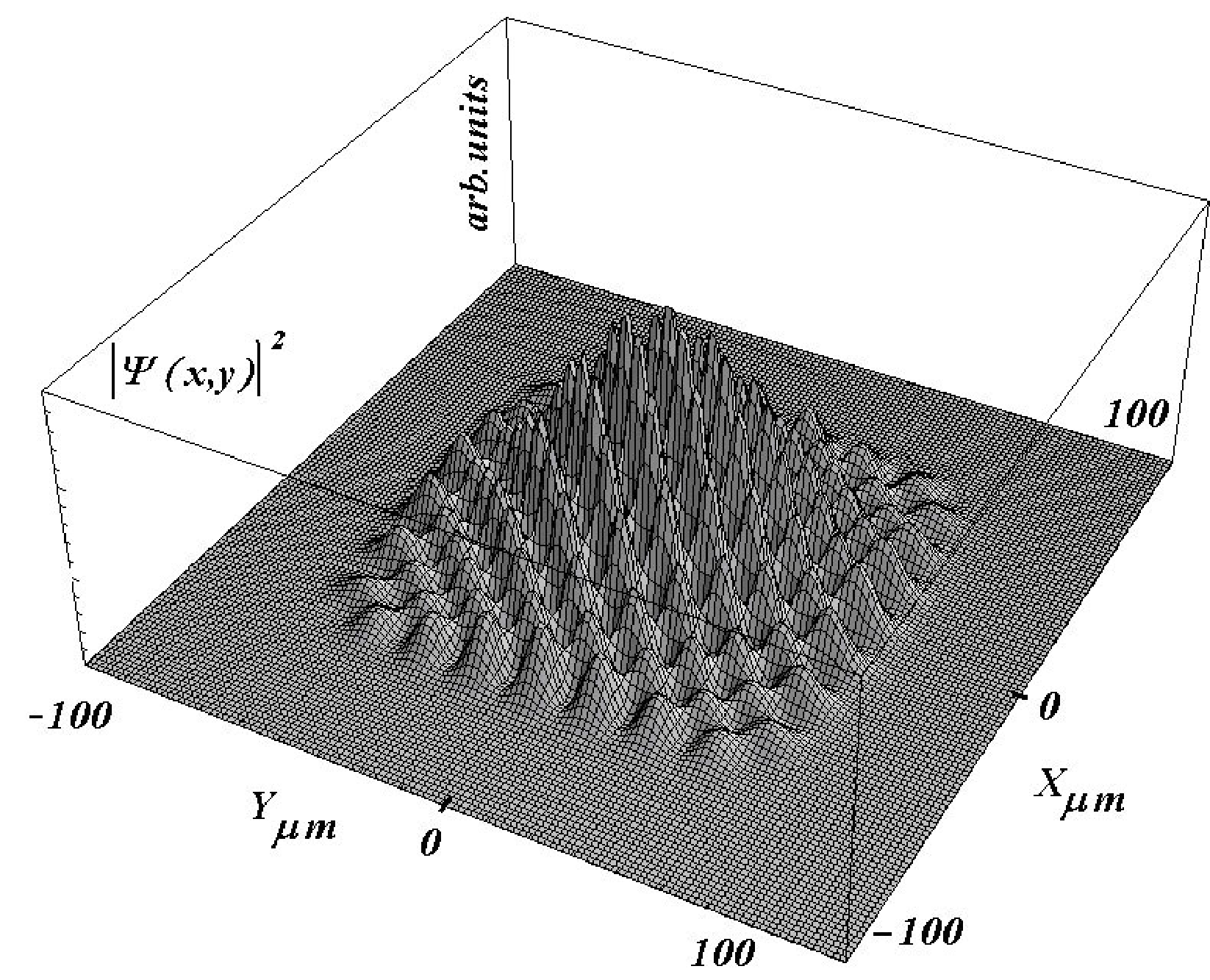}}
\caption{  The $\bf 2D$ plot of the square modulus
of the macroscopic wavefunction $\Psi ({\vec r},z=0)$
in the  $x,y$- plane. }
\label{fig.4}
\end{figure}

This condition could be fulfilled by virtue of the tuning scattering
length  $a_s$ via Feshbach resonance \cite{Pitaevskii:1999}:

\begin{equation}
\label{Feshbach}
  {a_s(\mathbf {|\vec{B}|})}=   { {{a}_s}^{bg}}{\:}
({1+\frac{{  {\Delta_{  {B}}}}}{  {B-B_{R}}}})
\end{equation}
where $  {\Delta_{  {B}}}$ is a width of Feshbach resonance,
 $  {B_{R}}$ is the value of the resonant magnetic field, ${ {{a}_s}^{bg}}$ is
background value of s-wave scattering length $a_s$.
The 3D solution $\Psi(\vec r, t)$ including numerical
evaluation of (\ref{GP2D})
corresponds to "pancake"-like BEC cloud aligned 
in the vicinity of  the $Z=0$ - plane (fig. \ref{fig.1}).
The superfluid vortices are collocated with the phase singularities
of the optical field. 
In contrast to the rotating "bucket" trap \cite{Ketterle:2002,Danaila:2005} 
and rotating "basket" trap 
\cite{Cornell:2006}, where superfluid vortex lattices rotate
as a rigid body, our solution (fig.  \ref{fig.4}) is static. 
The superfluid vortices in our static "basket" trap proved to be 
pinned at the nodes (i.e. zeros of amplitude or phase singularities) of the optical
interference pattern. The argument of $\Psi_{\bot}$
versus transverse coordinates shows
clearly the loci of rectangularily spaced vortices with
alternating circulations  (fig.  \ref{fig.5}).
The elementary superfluid vortices are labelled by white circles and squares.
The topological charges $\ell$ of the vortices labelled by
circles are $\ell =+1$  , whereas the closely neighbouring
squares have the opposite charges  $\ell =-1$.

\section{Rotational energy of the vortex-antivortex quantum state.}

Following to Feynman \cite{Feynman:1972} consider first the rotational
energy of an isolated vortex. In contrast to the vortex in a classical liquid which
rotates as a rigid body and have the constant angular velocity 
$ \omega(r)= v(r)/r $, where
$v(r)$ - is speed of flow line at the distance $r$ from vortex core, the quantum
liquid rotates in such a way that the phase $\theta $ of wavefunction $\Psi$ 
remains single-valued. Because the argument of the wave function $\theta $ is connected
with velocity $\vec v(\vec r)$ of superfluid via Madelung transform
$\nabla \theta(\vec r , t) = m {\vec v}/{\hbar}$, the contour integral
$\oint \nabla \theta(\vec r , t){\:}{d \vec l}= 2 \pi r m { v}/{\hbar} $
around the vortex core must be a multiple of $2 \pi$. As a consequence
the quantization of angular
 momentum follows $m v r = {\hbar} $ because of a single-valued phase
 $\theta $ of the wavefunction $\Psi_{\bot}$. The next step is in evaluation
of rotational
kinetic energy of the vortex using the classical definition:
$ \int\limits_{a}^{b}  {\omega}^2 (r) {\:} d J(r)/2 $.
The moment of inertia of an infinitesimally thin ($dr$)
ring $d J(r)={r^2}{\rho}{\:} 2 \pi r{\:} dr{\:} \chi$
rotating with angular velocity $\omega (r)= v/r = \hbar /(m r^2) $ is
integrated from the inner radius $a$ of the vortex core to the external one $b$:

\begin{eqnarray}
\label{rot_energ}
E_{rot} = {\int\limits_{a}^{b}}  {\omega}^2 
(r){\:}d{\:}J(r)/2=
{\chi}{\:}{\rho}{\int\limits_{a}^{b}} {\hbar}^2 
{\frac {\pi}{mr}} {\:}dr = 
& &  \nonumber \\
{\chi}{\:}{\hbar}^2 {\rho}{\:}{\frac {\pi}{m}}{\:}ln(b/a) ,
\end{eqnarray}
where $\chi$ is the length of vortex line in $z$-direction. Angular momentum of the vortex line
 may be determined by an analogous classical procedure:

\begin{figure}
\center{\includegraphics[width=0.8\linewidth] {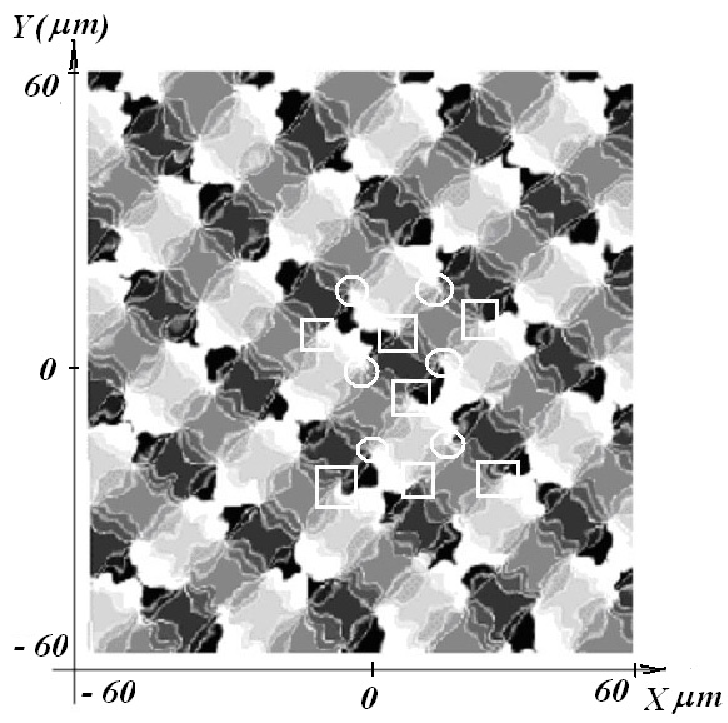}}
\caption{ The $\bf 2D$  plot of phase $\theta$ of macroscopic wavefunction
$\Psi ({\vec r},z=0)$ in the    $x,y$- plane.
The locations of the condensate vortices with positive topological
charges $\ell = +1$  are labeled by white circles.
The vortices having
the opposite charges  $\ell = -1$  are labeled by white squares.
$Z$ - axis is normal directed to reader.}
\label{fig.5}
\end{figure}

\begin{equation}
\label{angmom}
L_{vort} = {\int\limits_{a}^{b}}{\omega} 
(r){\:}d{\:} J(r) =
{\chi}{\:}{\pi}{\hbar} {\:}{\rho}
{\int\limits_{a}^{b}}{\frac {r{\:} dr}{m}} =
{\chi}{\:}{\pi}{\hbar} {\:}{\rho}{\:}
{\frac {b^2 - a^2}{2 {\:}m}},
\end{equation}

Vectorial nature of angular momentum ${\vec L }$ means that
for a rectangular
array of equispaced vortices with opposite circulations the local
${\vec L_{i,j} }$ are counter-directed (fig.). Thus the total
angular momentum of the array tends to be equal to zero:

\begin{equation}
\label{rot_ang_momentum}
{\vec L} = {\sum\limits_{jx,jy}}{\:}{\vec L_{jx,jy} }  \approx 0.
\end{equation}

On the other hand, the energies of the vortices are positive scalars hence their
energy in a rotational ground state are additive values for noninteracting vortices:

\begin{equation}
\label{rot_energ1}
 E_{ground} = {\sum\limits_{jx,jy}} E_{jx,jy} =
2 \times {{N}^2}_{vortices} \times {\chi}
{\:}{\hbar}^2 {\rho}
{\frac {\pi}{m}}{\:}\ln(b/a)
 {\:}{\:}{\:}{\:}{\:}{\:}
\end{equation}

The quantum mechanical evaluation of the energy and angular momentum
of vortex-antivortex quantum state is
performed as follows \cite{Reijnders:2005}. 
By definition of quantum expectation
values we have for the kinetic energy of 
the condensate \cite{Landau:1977}:

\begin{eqnarray}
\label{rot_energ_quant}
E_{ground}{\:} ={\:}<{\Psi}^{*}| \hat H | \Psi > =
& &  \nonumber \\
{\chi}{\int}{\int}{\Psi}^{*}
\lbrace - {\frac {\hbar^2 \Delta_{\bot}}{2m}}+ 
V(\vec r_{\bot})+ 
{\frac {4 \pi {\:}{\hbar}^2 a_s}{m}}
{|\Psi|}^2 \rbrace \Psi {\:} d^{{\:}2} {\vec r_{\bot}} 
& &  \nonumber \\
\approx {{N}^2}_{vortices} \times {\chi}{\:}{\hbar}^2 
{\rho}
{\frac {\pi}{m}}{\:} \ln (b/a),{\:}{\:}{\:}{\:}{\:}{\:}
\end{eqnarray}

and for the angular momentum $L_z$:

\begin{equation}
\label{angular_momentum1}
{\:}{\:}  L_{z}{\:} = {\:}<{\Psi}^{*} | \hat L | \Psi>{\:} =  {\:}
<{\Psi}^{*} | -i \hbar {\frac {\partial}{\partial \phi}} | \Psi> \approx 0.
\end{equation}

\section{Conclusion.}
The  outlined optical labyrinth trap setup
is capable to support an asymmetrical cloud of
 ultracold atoms "red" detuned from resonance.
The macroscopic quantum state obtained under facrorization conditions:
the longitudinal component is $z$ - dependent gaussian with
characteristic width of harmonic oscillator's ground state,
while transversal part is periodic rectangular vortex lattice,
pinned by the vortices of trapping optical lattice.
The optical torque
will be shown to cause the atoms roaming across optical lattice.


\end{document}